\documentclass[10pt,conference]{IEEEtran}

\usepackage{amsmath,amssymb,amsfonts}
\usepackage{algorithmic}
\usepackage{graphicx}
\usepackage{textcomp}
\graphicspath{{./figs/}}
\usepackage{cite}
\usepackage[latin1]{inputenc}
\usepackage{color}
\usepackage{multirow}
\usepackage{siunitx}

\usepackage{url}

\def\BibTeX{{\rm B\kern-.05em{\sc i\kern-.025em b}\kern-.08em
		T\kern-.1667em\lower.7ex\hbox{E}\kern-.125emX}}

\definecolor{light-gray}{rgb}{0.8,0.8,0.8}
\definecolor{BrickRed}{rgb}{0.8,0.0,0.0}
\definecolor{Brown}{rgb}{0.4,0.2,0.2}

\makeatletter
\newcommand{\newlineauthors}{%
\end{@IEEEauthorhalign}\hfill\mbox{}\par
\mbox{}\hfill\begin{@IEEEauthorhalign}
}
\makeatother

\newcommand{\norm}[1]{\left\lVert{#1}\right\rVert}

\newcommand{\argmax}[1]{\underset{#1}{\operatorname{argmax} \ }}

\def\bsalpha{{\boldsymbol{\alpha}}}

\def\bseta{{\boldsymbol{\eta}}}

\def\bssigma{{\boldsymbol{\sigma}}}

\def\bsp{{\mathbf{p}}}

\def\bsS{{\mathbf{S}}}

\def\nr{{N_\text{r}}}

\begin{document}
\title{Using Mobile Phones for Participatory Detection and Localization of a 
	GNSS Jammer
\thanks{This work was supported in part by Security Link and SSF-SURPRISE}
}

\author{\IEEEauthorblockN{Gl\"adje Karl Olsson}
	\IEEEauthorblockA{\textit{Dept. for Sensor Networks and Multi Sensor 
	Fusion} 
\\
		\textit{Swedish Defence Research Agency}\\
		Link\"oping, Sweden \\
		gladje.karl.olsson@foi.se}
	\and
	\IEEEauthorblockN{Sara Nilsson}
	\IEEEauthorblockA{\textit{Dept. for EW Radio Countermeasures} 
\\
		\textit{Swedish Defence Research Agency}\\
		Link\"oping, Sweden \\
		sara.nilsson@foi.se}
%	\and
	\newlineauthors
	\IEEEauthorblockN{Erik Axell}
	\IEEEauthorblockA{\textit{Dept. for Robust Radio Communications} 
\\
		\textit{Swedish Defence Research Agency}\\
		Link\"oping, Sweden \\
		erik.axell@foi.se}
	\and
	\IEEEauthorblockN{Erik G. Larsson}
	\IEEEauthorblockA{\textit{Dept. of Electrical Engineering (ISY)} 
\\
		\textit{Link\"oping University}\\
		Link\"oping, Sweden \\
		erik.g.larsson@liu.se}
	\and
	\IEEEauthorblockN{Panos Papadimitratos}
	\IEEEauthorblockA{\textit{Networked Systems Security Group} 
\\
		\textit{Royal Institute of Technology (KTH)}\\
		Stockholm, Sweden \\
		papadim@kth.se}
}

\IEEEoverridecommandlockouts

%% For arXiv publishing
\IEEEpubid{\begin{minipage}{\textwidth}
\copyright2023 IEEE. Personal use of this material is permitted. Permission 
from \\
IEEE must be obtained for all other uses, in any current or future media, \\
including reprinting/republishing this material for advertising or promotional 
\\
purposes, creating new collective works, for resale or redistribution to 
servers \\
or lists, or reuse of any copyrighted component of this work in other works.
\end{minipage}}

\maketitle

\IEEEpubidadjcol

\begin{abstract}
It is well known that GNSS receivers are vulnerable to jamming and spoofing
attacks, and numerous such incidents have been reported in the last decade all
over the world. The notion of participatory sensing, or crowdsensing, is that a
large ensemble of voluntary contributors provides measurements, rather than
relying on a dedicated sensing infrastructure. The participatory sensing network
under consideration in this work is based on GNSS receivers embedded in, for
example, mobile phones. The provided measurements refer to the receiver-reported
carrier-to-noise-density ratio ($C/N_0$) estimates or automatic gain control
(AGC) values. In this work, we exploit $C/N_0$ measurements to locate a GNSS 
jammer, using multiple receivers in a crowdsourcing manner. We extend a previous
jammer position estimator by only including data that is received during parts 
of the sensing period where jamming is detected by the sensor. In addition, we 
perform hardware testing for verification and evaluation of the proposed and 
compared state-of-the-art algorithms. Evaluations are performed using a Samsung 
S20+ mobile phone as participatory sensor and a Spirent GSS9000 GNSS simulator 
to generate GNSS and jamming signals. The proposed algorithm is shown to work 
well when using $C/N_0$ measurements and outperform the alternative algorithms 
in the evaluated scenarios, producing a median error of 50 meters when the 
pathloss exponent is 2. With higher pathloss exponents the error gets higher. 
The AGC output from the phone was too noisy and needs further processing to be 
useful for position estimation.
\end{abstract}


\section{Introduction}
\label{sec:introduction}

Global Navigation Satellite System (GNSS) receivers are widely spread in 
society, including society-critical services, today. It is also known that GNSS 
receivers are vulnerable to jamming and spoofing attacks, and numerous such 
incidents have been reported in the last decade all over the world. Therefore, 
it is important to detect and localize the source of such attack, which can be 
hard without a dedicated sensing infrastructure. The notion of participatory 
sensing, or crowdsensing, is that a large ensemble of voluntary contributors 
provides the measurements, rather than relying on a dedicated sensing 
infrastructure.

The participatory sensing network under consideration in this work is based on 
GNSS receivers embedded in connected devices, for example mobile phones. The 
provided measurements refer to the receiver-reported carrier-to-noise-density 
ratio ($C/N_0$) estimates or automatic gain control (AGC) values. $C/N_0$ 
measurements are provided by all grades of GNSS receivers, from low-cost to 
professional. Some embedded receivers, for example modern Android devices 
(cf. \cite{Spens-NAVIGATION-2022}), also output AGC gain values. The AGC is 
used to maintain a desired signal amplitude at the receiver, even though the 
incoming signals vary in amplitude. Both of these measurements allow for 
detection and localization of the source of jamming using different types of 
embedded GNSS receivers. Similar ideas have been considered in previous work on 
crowdsensing-based GNSS interference detection and localization 
(cf. \cite{Olsson-ICLGNSS-2022, Strizic-ITM-2018, 
	Kraemer-GNSS-2012, Borio-GNSS+-2016, Ahmed-ICL-GNSS-2020, 
	Ahmed-AeroConf-2021, 
	Liu-CSNC-2020, Wang-TAES-2020}). These papers exploit jamming power 
estimation, via $C/N_0$ estimates, AGC values, or via direct power measurements.

Crowdsourcing of data from mobile phones for GNSS jamming detection and 
localization was considered in \cite{Strizic-ITM-2018, Kraemer-GNSS-2012}. Field
trials showed that AGC or $C/N_0$ measurements from mobile phones could be used 
for jammer localization \cite{Strizic-ITM-2018}. There is, however, no 
localization algorithm explicitly proposed in \cite{Strizic-ITM-2018}. A 
localization algorithm based on the combination of $C/N_0$ measurements, step 
detection and step length estimation using an inertial sensor was proposed in 
\cite{Kraemer-GNSS-2012}. However, the algorithm of \cite{Kraemer-GNSS-2012} 
assumes that the receiver moves along a straight line in two perpendicular 
directions. Such restrictions on the receiver movement are not generally 
applicable, as they may be hard to implement for a participatory sensing 
scenario. It is desirable to have the users act freely, as desired and 
seamlessly collect and contribute data.

The system model assumptions in this work are the same as in 
\cite{Olsson-ICLGNSS-2022} and very similar to the models used in 
\cite{Borio-GNSS+-2016, Ahmed-ICL-GNSS-2020, Ahmed-AeroConf-2021, 
	Wang-TAES-2020, Liu-CSNC-2020, Tucker-GNSS+-2020}. The main differences are
that our work and \cite{Olsson-ICLGNSS-2022} extend the models of 
\cite{Borio-GNSS+-2016, Ahmed-ICL-GNSS-2020, Ahmed-AeroConf-2021, 
	Wang-TAES-2020, Liu-CSNC-2020, Tucker-GNSS+-2020} by including a stochastic 
measurement error and assuming, more realistically, that the path loss between 
each receiver and the jammer is unknown. In addition, the algorithm proposed in
\cite{Olsson-ICLGNSS-2022}, and further extended in this work, does not require 
calibration of the receivers with respect to the jamming power. This is 
required by the algorithms of \cite{Borio-GNSS+-2016, 	Ahmed-ICL-GNSS-2020, 
	Ahmed-AeroConf-2021, Wang-TAES-2020}.

$C/N_0$-based power estimation and a distance-dependent pathloss model were 
used in \cite{Borio-GNSS+-2016, Ahmed-ICL-GNSS-2020, Ahmed-AeroConf-2021, 
	Liu-CSNC-2020} to compute a least-squares (LS) estimate of the jammer 
position. The algorithm of \cite{Borio-GNSS+-2016} was derived for a single 
receiver that moves and thereby can be viewed as a synthetic array. Another
algorithm was also proposed in \cite{Borio-GNSS+-2016} that uses data from 
multiple receivers and uses the average location of all those receivers that
detect the jammer. That algorithm was extended in \cite{Wang-TAES-2020} to 
obtain a weighted average location estimate, instead of the ordinary arithmetic 
mean.

Reference \cite{Tucker-GNSS+-2020} considered a power difference of arrival 
(PDOA) algorithm, which is similar to exploiting $C/N_0$ or AGC measurements. 
However, the algorithm of \cite{Tucker-GNSS+-2020} measures the received power
directly and is therefore not suited for application to embedded receivers, 
which do not use such data in general.

Our work considers GNSS receivers embedded in mobile phones, to provide 
automatic gain control (AGC) or $C/N_0$ estimates for participatory jamming 
detection and localization. The proposed algorithm does not require knowledge 
of the jamming transmit power nor of the path loss between the jammer and each 
sensor, but automatically estimates all parameters.

More specifically, we here extend the work of \cite{Olsson-ICLGNSS-2022} by
\begin{enumerate}
	\iffalse
	\item Extending the algorithm of \cite{Olsson-ICLGNSS-2022} to include data 
	for jammer position estimation that is received during only parts of the 
	sensing period. In \cite{Olsson-ICLGNSS-2022} the whole sequence 
		was used if a jamming was detected. Now, only the samples when the 
		receiver is jammed is used. 
	\fi
	\item Extending the algorithm of \cite{Olsson-ICLGNSS-2022} to only include
	received data during times when jamming is detected. In 
	\cite{Olsson-ICLGNSS-2022} the whole sequence was used if an attack was 
	detected, and thus might include data without any jamming which can affect 
	the estimate in a bad way.
	\item Hardware testing for verification and evaluation of the proposed and 
	compared state-of-the-art algorithms of \cite{Borio-GNSS+-2016, 
		Ahmed-ICL-GNSS-2020}. Evaluations are performed using a Samsung S20+ 
	mobile phone as participatory sensor and a Spirent GSS9000 GNSS simulator to 
	generate GNSS and jamming signals. The proposed algorithm is shown to work 
	well and outperform the compared algorithms in the evaluated scenarios.
\end{enumerate}

The paper is organized as follows: Section~\ref{sec:model} presents the 
assumptions and system model. Section~\ref{sec:proposed} briefly explains the 
algorithm of \cite{Olsson-ICLGNSS-2022} and the proposed extensions to 
it. The hardware system setup and the results of the evaluation are shown in  
Section~\ref{sec:simulation_setup} and Section~\ref{sec:results} respectively. 
Finally, Section~\ref{sec:conclusion} concludes the paper.
\section{System Model}
\label{sec:model}

The system model is adopted from \cite{Olsson-ICLGNSS-2022} and briefly 
described here for completeness of the exposition. The following assumptions are
made:
\begin{itemize}
	\item Receive and transmit antennas are isotropic.
	\item The jammer position, denoted by $\bsp_0 = [x_0, y_0, z_0]^T$, is fixed
	during the duration of the measurement.	
	\item The unknown jamming power, $J_0$, is constant during the duration of 
	the measurement.
	\item The receiver noise powers are constant during the duration of the 
	measurement.
	\item The receiver positions at each time instant are known, and denoted by 
	$\bsp_i[n]=[x_i[n],y_i[n],z_i[n]]^T$ at time $n$ for receiver 
	$i=1,\ldots,N_r$.
\end{itemize}

The goal is to estimate the jammer position, $\bsp_0$, using $C/N_0$-estimates 
taken from commercial mobile phones. What the receivers estimate, which is 
usually referred to as $C/N_0$ with slight abuse of terminology, is actually the
carrier-to-noise-and-interference-density (CNIR). To avoid this abuse of 
terminology, the measured metric is from here on referred to as CNIR.

With $C_{i,j}[n]$ denoting the received power at receiver $i$ from satellite $j$
at time $n$, the background noise power spectral density $N_{0,i}$, and 
$\tilde J_i[n]$ as the received spectrally-flat-equivalent interference power 
density, the CNIR can be expressed as:
\begin{equation*}
	s_{i,j}[n] \triangleq \frac{C_{i,j}[n]}{N_{0,i}+\tilde J_i[n]} = 
	\frac{C_{i,j}[n]}{N_{0,i}}\cdot \frac{1}{1+\frac{\tilde J_i[n]}{N_{0,i}}}.
\end{equation*}

Canceling the bandwidth dependency by multiplying $\tilde J_i[n]$ and $N_{0,i}$ 
with the receiver bandwidth, and taking the mean CNIR values from all satellite 
signals results in 
\begin{equation*}
	s_i[n] \triangleq 
	\frac{1}{N_m}\sum_{j=1}^{N_m}\left(\frac{C_{i,j}[n]}{N_{0,i}}\cdot 
	\frac{1}{1+\frac{J_i[n]}{N_{S,i}}}\right),
\end{equation*}
where $N_{S,i}$ and $J_i[n]$ are, at receiver $i$, the noise power in the CNIR 
estimation and the received jamming power at time $n$, respectively, and $N_{m}$
is the number of satelites. 

Let $\bar s_i$ be the mean CNIR when no jamming is present ($J_i[n] = 0$), 
which can be estimated during initialization. With
\begin{equation*}
	\bar s_i = \frac{1}{N_m}\sum_{j=1}^{N_m}\frac{C_{i,j}[n]}{N_{0,i}},
\end{equation*}
the expression for $s_i[n]$ can be written as
\begin{equation}
	\label{eq:cnir_linear_measurement}
	s_i[n] = \bar s_i\cdot \frac{1}{1+\frac{J_i[n]}{N_{S,i}}}.
\end{equation}

By adopting a simple distance-dependent path loss model, the received jamming 
power can be written
\begin{equation*}
	J_i[n] = \frac{J_0 \kappa}{d_i[n]^{\alpha_i}},
\end{equation*}
where $\alpha_i$ is the path loss exponent for 
receiver $i$, the distance $d_i[n] \triangleq \norm{\bsp_0 - \bsp_i[n]}$ and 
$\kappa$ is a proportionality constant.

Rewriting the expression \eqref{eq:cnir_linear_measurement} in logarithmic 
scale, with $S_i[n]$ and $\bar S_i$ denoting $s_i[n]$ and $\bar s_i$ in 
decibels, the measurement model for CNIR is
\begin{equation*}
	\label{eq:Cn0_meas_model}
	S_i[n] = \bar S_i - 
	10\log_{10}\left(\frac{J_0\kappa}{N_{S,i}}d_i^{-\alpha_{i}}[n]+1\right) + 
	w_{S_i}[n],
\end{equation*}
where $w_{S,i}[n]\sim\mathcal{N}(0,\sigma^2_{S_i})$ is additive zero-mean white 
Gaussian noise. 

The AGC-values from the receiver can also be used in a similar manner as the 
CNIR measurements, and results in an analogous measurement model. This is 
described more thoroughly in \cite{Olsson-ICLGNSS-2022}.

\section{Proposed Algorithm}
\label{sec:proposed}

To locate the transmitter, the likelihood function is maximized using gradient 
descent on the negative log-likelihood. Let the individual variables for all 
receivers, $i = 1, \ldots, \nr$, be collected in vectors denoted by a boldface
equivalent, in analogy with $\bsS \triangleq [S_1, \ldots, S_\nr]^T$. With 
$\bsS$ as the received CNIR measurements from the sensors, the likelihood 
function is given by 
\begin{equation}
	\label{eq:likelihood_func}
	\begin{split}
		&p(\bsS|\bar \bsS, \bsp_0, \bsalpha, \bseta, \bssigma^{2}_{S}) =\\
		&= \prod_{i=1}^{Nr}(2\pi\sigma^{2}_{S,i})^{-N/2}\exp\Big[-\frac{1}
		{2\sigma^{2}_{S,i}}\sum_{n=1}^{N}(S_{i}[n]- \\
		&(\bar S_{i}-10\log_{10}(\eta_{i}d^{-\alpha_{i}}_{i}[n]+1)))^{2}\Big],
	\end{split}
\end{equation}
 where $\eta_{i} \triangleq \frac{J_0\kappa}{N_{S,i}}$. The sought transmitter 
 position is estimated as 
 \begin{equation*}
 	\bsp_0^{*} = \argmax{\bsp_0,\bseta,\bssigma^{2}_{S},\bsalpha} p(\bsS|\bar 
 	\bsS,\bsp_0,\bsalpha,\bseta,\bssigma^{2}_{S}).
 \end{equation*}

The unknown pathloss is not estimated during the gradient descent, but it is 
instead estimated using a grid search for each individual sensor. Receivers that
have a free line-of-sight to the jammer, i.e. $\alpha_{i}\approx2$, are assumed
to adhere to the pathloss model and are therefore deemed to be more efficient in
estimating the distance to the jammer. Therefore, those sensors that gets an
estimated $\alpha_{i}\leq2.3$ are selected for joint estimation of the jammer 
position by maximizing the likelihood function once again, but sensors that 
estimate $\alpha_{i} > 2.3$ are excluded from the estimation.

The maximization of the likelihood function in Equation 
\eqref{eq:likelihood_func} cannot be done analytically, which is why a gradient
descent is performed. However, the variable $\bssigma^{2}_{S}$ can be solved
as function of the other unknowns, as 
\begin{equation*}
	\sigma^{2}_{S,i} = \frac{1}{N}\sum_{n=1}^{N}(S_{i}[n] - \bar S_{i} + 
	10\log_{10}(\eta_{i}d^{-\alpha_{i}}_{i}[n]+1))^{2}.
\end{equation*}
This estimate of $\bssigma^{2}_{S}$ is updated after each iteration of the
gradient descent.

One extension that is made from \cite{Olsson-ICLGNSS-2022} is that the algorithm
now only includes data at time instants when the receiver is detecting an
attack, i.e. when $S_{i}[n]-\bar S_{i}$ falls below a predetermined decision
threshold. This is done to ensure that the used samples are contributing relevant
information. A sample without notable jamming present is not giving much 
information on where the jammer is located. This extensioncan be formulated by
rewriting Equation \eqref{eq:likelihood_func} as:
\begin{equation*}
	\begin{split}
		&p(\bsS|\bar \bsS, \bsp_0, \bsalpha, \bseta, \bssigma^{2}_{S}) =\\
		&= \prod_{i=1}^{Nr}(2\pi\sigma^{2}_{S,i})^{-N/2}\exp\left[-\frac{1}{2\sigma^{2}_{S,i}}\sum_{n=1}^{N}(X_i[n])^{2}\right],
	\end{split}
\end{equation*}
where $X_i[n]$ is the conditional variable
\begin{equation*}
	X_i[n] =
		\begin{cases}
			\begin{aligned}
				&S_{i}[n]-\bar S_{i} + \\ 
				& \ \ 10\log_{10}(\eta_{i}d^{-\alpha_{i}}_{i}[n]+1), 
			\end{aligned} & \text{if}\ S_{i}[n]-\bar S_{i} < \gamma, \\
			0, & \text{otherwise},
		\end{cases}
\end{equation*}
and $\gamma$ is the decision treshold. When no attack is detected at time $n$, 
the term $X_i[n]$ is set to $0$ for that $n$, and it is then not contributing 
to the likelihood function. This is also done if a receiver is totally saturated
by the jammer, such that data is not output by the receiver at certain time 
instants.
\section{Hardware Simulation Setup}
\label{sec:simulation_setup}

\begin{figure}[t]
	\centering
	\includegraphics[width=1.0\columnwidth]{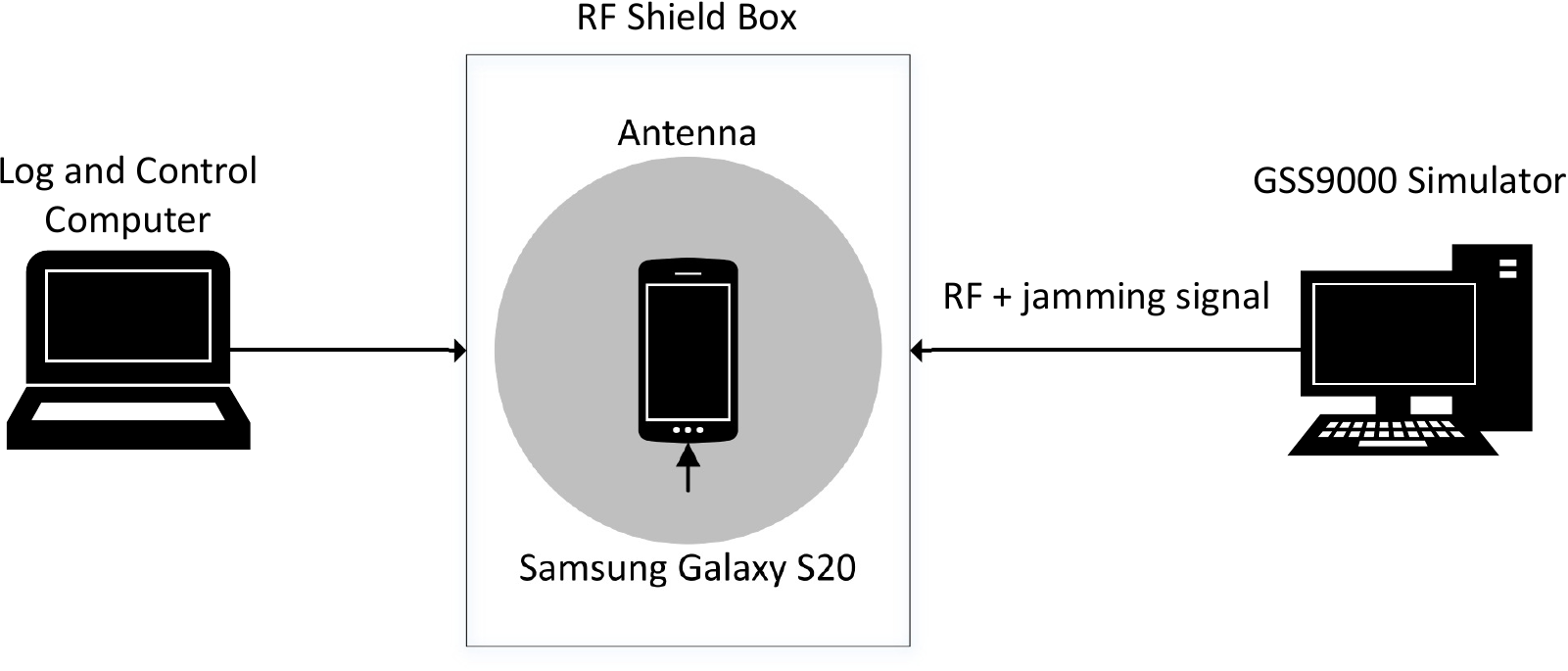}
	\caption{Hardware simulation setup.}
	\label{fig:hardware_system_setup}
\end{figure}

The hardware simulations were done using a Spirent GSS9000 GNSS simulator with
a GTx (i.e. jammer) option that generated the GNSS and jamming signals, and a 
Samsung Galaxy S20+ as the sensor. Since the phone has an internal antenna, the
GNSS and jamming signals have to be transmitted over the air. Therefore, the 
phone was placed in a universal RF shield box from Rhode \& Schwartz to receive
only the wanted signals. AGC and CNIR measurements were extracted from the phone
using the application GnssLogger \cite{GNSS-logger} for Android devices. The 
hardware simulation setup is shown in Figure \ref{fig:hardware_system_setup}. 

The GNSS signals were generated in the simulations to represent predefined 
positions and time that the phones would experience in a real world scenario at
these positions and time instants. Only GPS L1 signals were used. The phone had
to be set in flight mode and the time manually selected to be close to the start
time of the simulated scenario, for the phone to lock on to the simulated GPS 
signals.

The simulated movements of the phones during the scenario were generated as 
follows. After a start-up period of five minutes to get a position fix, the 
mobile phone moves along a straight line with constant velocity (80 km/h) and
the jamming signal is turned on. Numerous different straight line movements, 
each representing an individual phone, were generated to represent different 
receivers that are used for the participatory sensing. The jammer was located at
a fixed position during all simulations. These movements were designed to be 
similar to those evaluated in the software simulations in 
\cite{Olsson-ICLGNSS-2022}. The jamming signal was defined as a continuous wave 
centered at the GPS L1 frequency. The received jamming signal power is defined 
such that the path loss of the jamming signal matches the assumed system model. 
Different path loss exponents, $\alpha$, were used for the hardware simulations. 
The path loss values $\alpha$ = 2, 2.5 and 2.9338 are used for all movements. 
Because of the limited dynamic range of the jamming power in the GSS9000, the
jamming signal power was set to reach a minimum of 15 dB higher than the power
of each GNSS signal.

\begin{figure}[t]
	\centering
	\includegraphics[width=1.0\columnwidth]{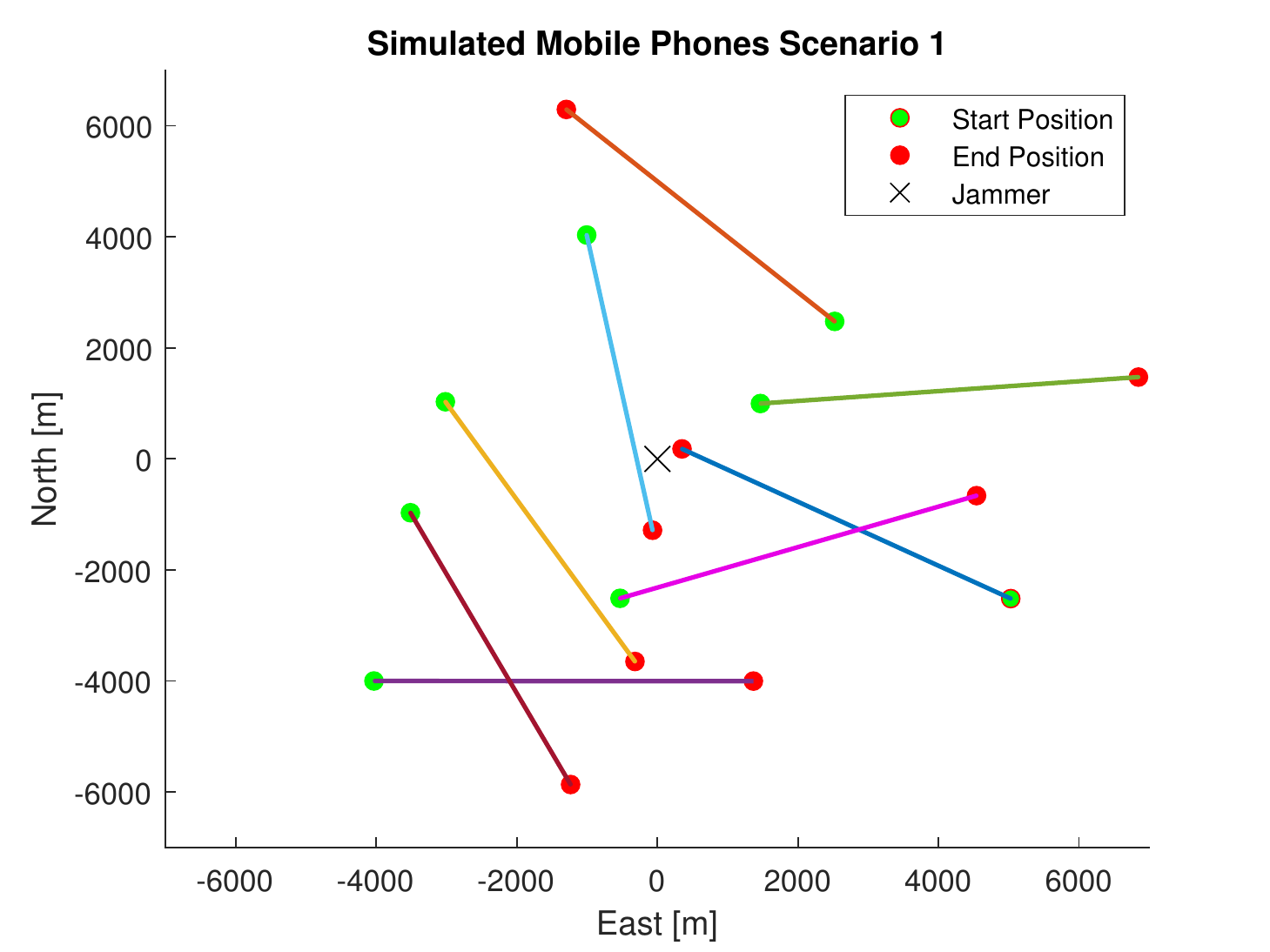}
	\caption{Jammer position and an example of receiver movements in hardware 
		evaluation for scenario 1.}
	\label{fig:extended_abstract_mobile_pos_scenario1}
\end{figure}

\section{Numerical Results}
\label{sec:results}
The jammer position is estimated by maximizing the proposed likelihood function 
in Equation \eqref{eq:likelihood_func} numerically. This is compared to two 
other methods, presented in \cite{Borio-GNSS+-2016} and 
\cite{Ahmed-ICL-GNSS-2020}. For jammer localization, the method of 
\cite{Borio-GNSS+-2016} estimates the jammer position as the mean position of 
all receivers that detect the jamming, while the algorithm in 
\cite{Ahmed-ICL-GNSS-2020} is based on a least-squares estimation method. The 
second method requires a calibration of the receivers with a known jamming 
power at a known distance. It also assumes the pathloss to be $\alpha_{i} = 2$
for all receivers, which is not very realistic. In reality, each receiver will 
be affected differently by multipath or faiding of the signal, which will result
in variations of the pathloss. In this evaluation, the proposed algorithm is 
tested both when $\alpha_i$ is assumed to be known and when it is estimated 
using a grid search. This is done to make a fair comparison to
\cite{Ahmed-ICL-GNSS-2020} that assumes perfect knowledge of $\alpha_{i} = 2$. 
Different values of $\alpha_i$ are also tested.

\subsection{Different Number of Sensors}
\label{sec:no_receivers}
\begin{figure}[t]
	\centering
	\includegraphics[width=1.0\columnwidth]{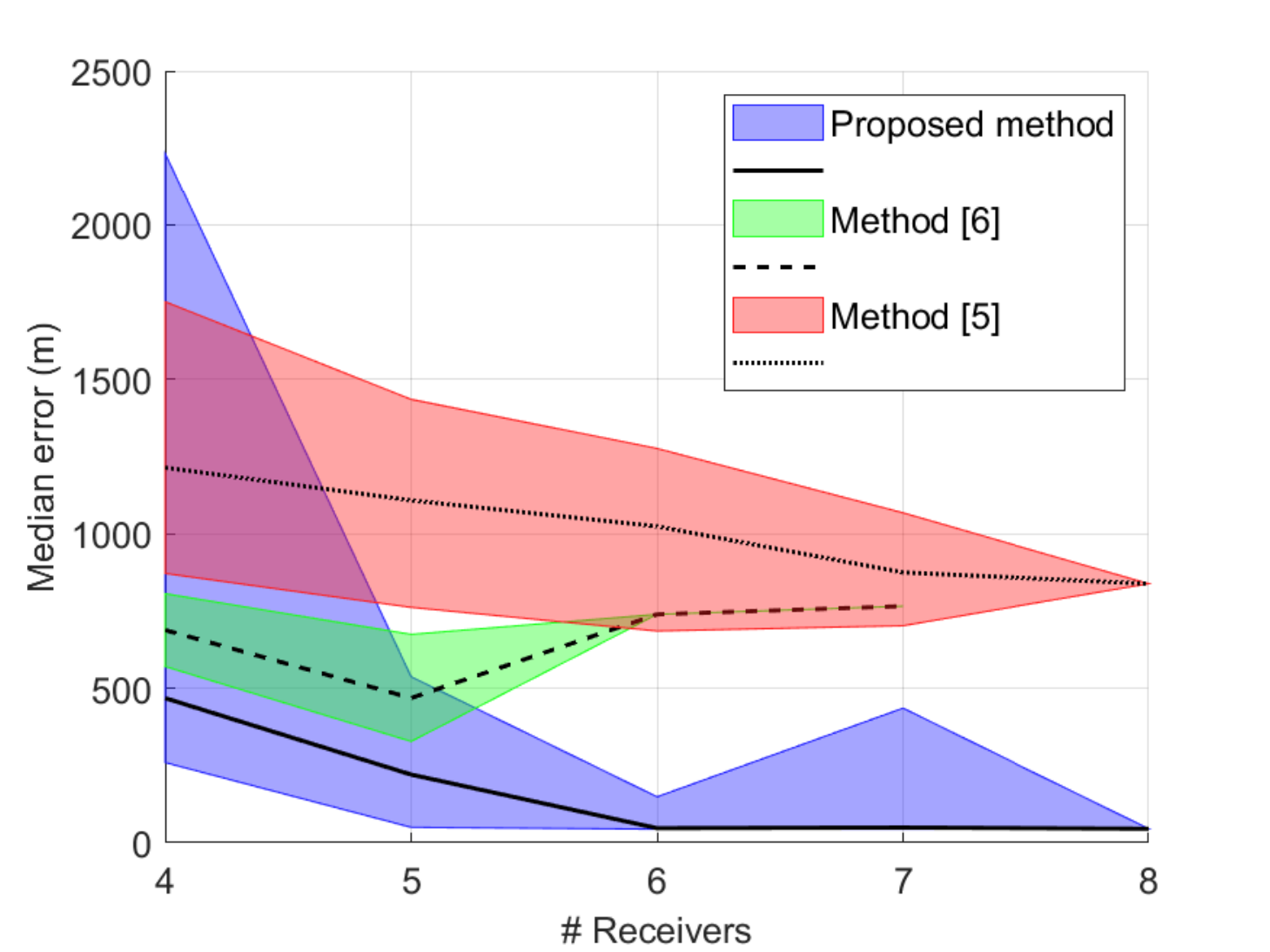}
	\caption{Median 3D-error (black lines) and 25th to 75th percentile (shaded
		areas) for the three different methods using different number of receivers.}
	\label{fig:alpha2_cluster}
\end{figure}

\begin{figure}[t]
	\centering
	\includegraphics[width=1.0\columnwidth]{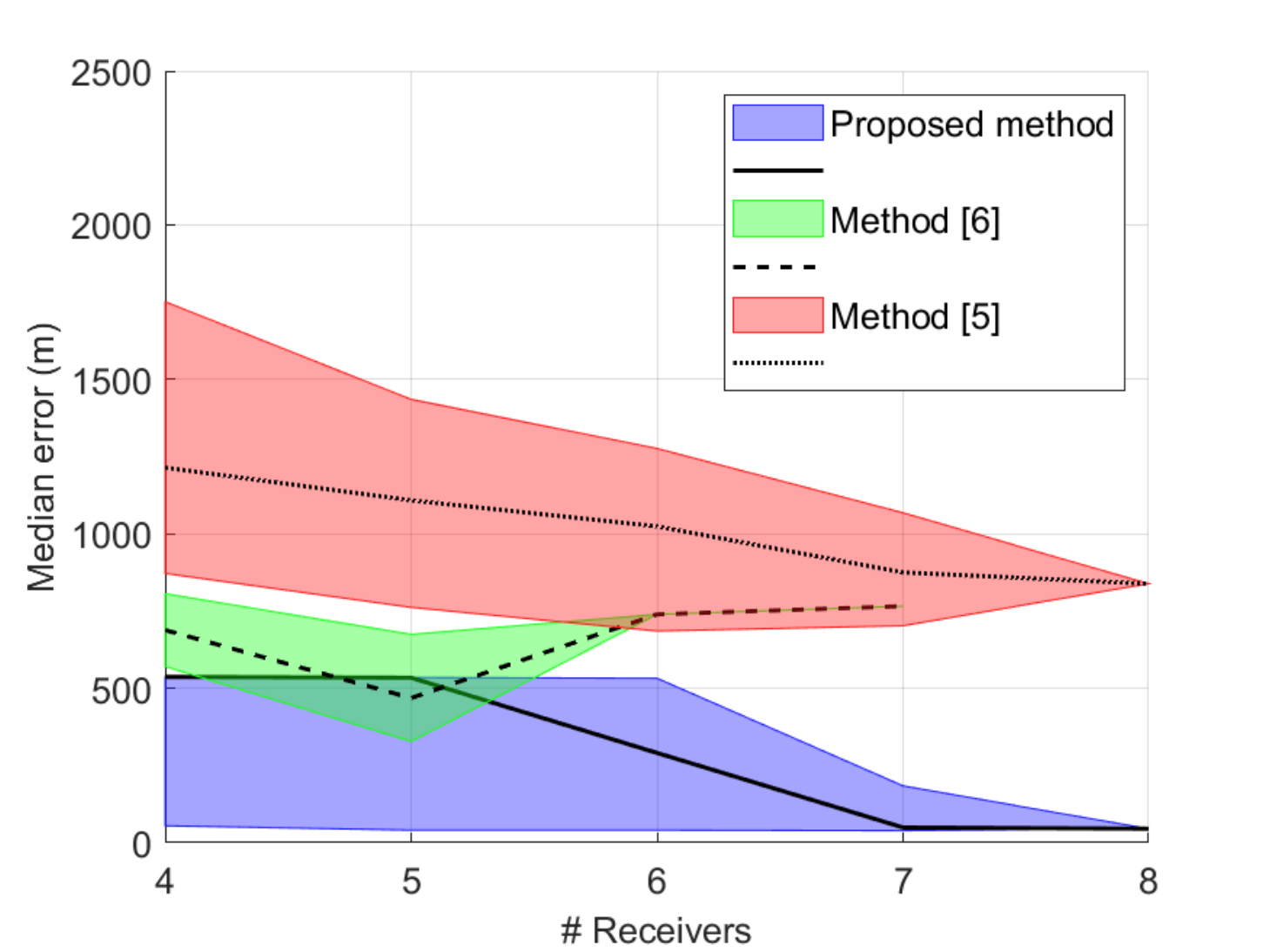}
	\caption{Median 3D-error (black lines) and 25th to 75th percentile (shaded
		areas) for the three different methods using different number of receivers.}
	\label{fig:alpha2_given}
\end{figure}

The algorithms are evaluated by testing different numbers of receivers. From the
eight possible receiver paths, all combinations using at least four different 
receivers are tested. The result is shown in Figure \ref{fig:alpha2_cluster} as 
the median position error as a function of the number of receivers. Figure 
\ref{fig:alpha2_given} shows results when the pathloss is perfectly known 
$(\alpha=2)$ and therefore the grid search is not utilized to estimate 
$\alpha_i$. For that reason, the exclusion of sensors with $\alpha_i > 2.3$ is
not applied in this case. With eight receivers, the number of receiver 
combinations are 70, 56,  28, 8 and 1, using 4 up to all receivers respectively. 

The method of \cite{Ahmed-ICL-GNSS-2020} has a hard time finding a solution at 
times, not solving the least squares-problem at all when using all eight 
receivers. The proposed method performs best among all three methods, where the 
median position error reaches down to around 50 meters using six or more 
receivers, while method \cite{Borio-GNSS+-2016} does not get better than 900 
meters, and method \cite{Ahmed-ICL-GNSS-2020} gets an error of about 500 meters 
at best.

When $\alpha$ is given, and the receiver exclusion step is skipped, the proposed
method gives slightly different results, shown in Figure \ref{fig:alpha2_given}.
At its best, the estimation performance is better, especially using only a few 
receivers. However, the median error of the proposed algorithm with known 
$\alpha_i = 2$ and no receiver selection in Figure~\ref{fig:alpha2_given} is 
noticeably worse than with estimated $\alpha_i$ and receiver selection in Figure
\ref{fig:alpha2_cluster}, with fewer than 7 receivers. Thus, the receiver 
selection might help filtering out the receivers that are most useful, and 
corresponds more with the measurement model. Another improvement due to the 
receiver selection is the number of times the algorithm finds a solution. 
Without exclusion of receivers, the algorithm does not converge 60-70\% of the
time, when using 4-6 receivers. With receiver selection, the failure to converge
drops to 20-25\%. Such convergence ratio is not desired, but it is still much 
better than method \cite{Borio-GNSS+-2016}, which does not converge 90\% of the 
time. In a future work, methods for a better convergence will be tested.


\subsection{Varying Pathloss Values}
\begin{figure}[t]
	\centering
	\includegraphics[width=1.0\columnwidth]{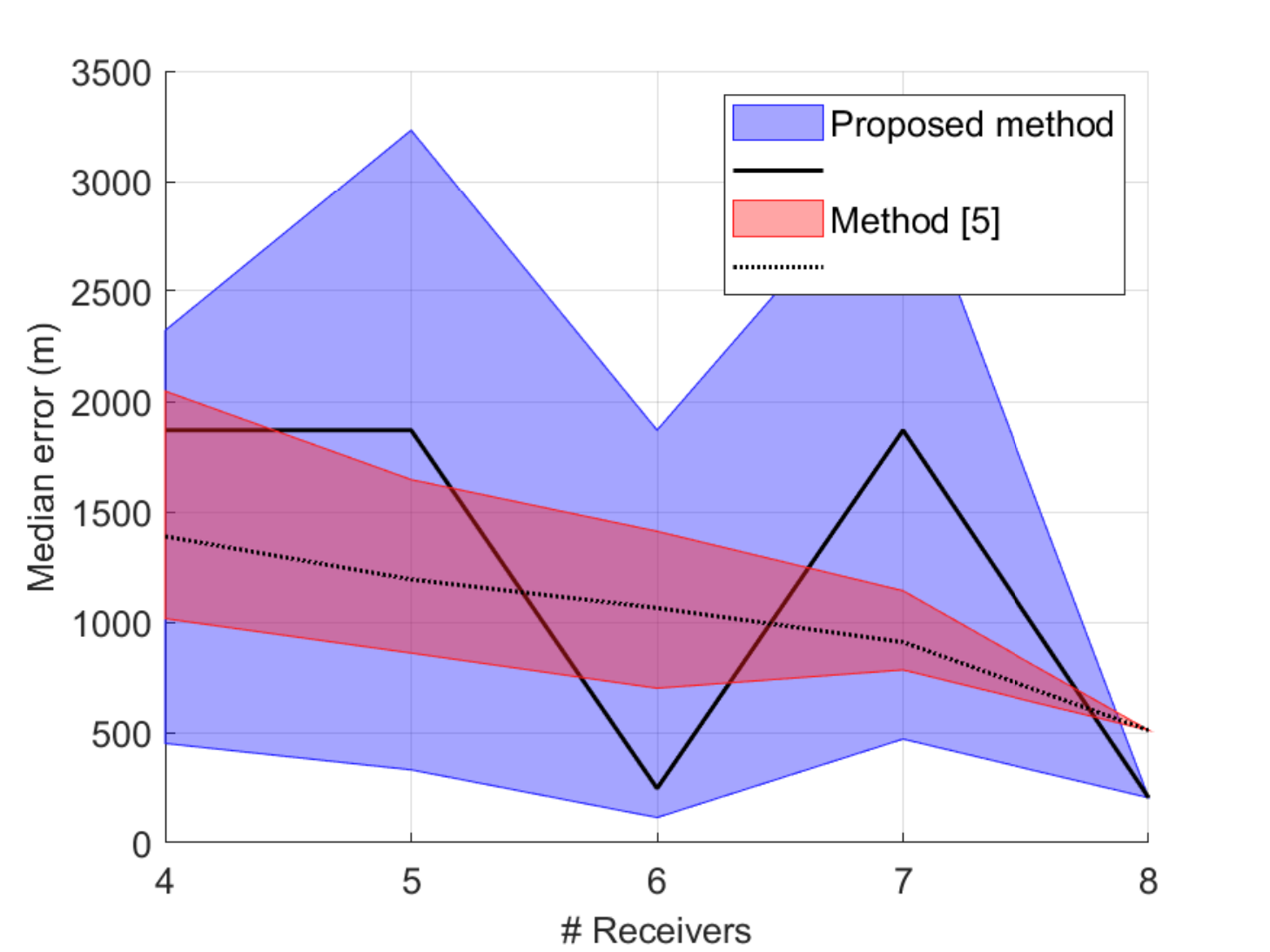}
	\caption{Median 3D-error (black lines) and 25th to 75th percentile (shaded
		areas) for the two different methods using different number of receivers.}
	\label{fig:varying_alpha}
\end{figure}

Different values of $\alpha$ are tested for the different receiver paths. The 
same test as in Section~\ref{sec:no_receivers} is excecuted, but the pathloss 
for each receiver is randomly set to one of the values $2, 2.5$ or $2.9338$,
which can differ between the receivers. The results are presented in Figure
\ref{fig:varying_alpha}. The method of \cite{Ahmed-ICL-GNSS-2020} is not 
included in the figure, because the least-squares problem could not be solved 
at all when the pathloss values are larger than 2.

The proposed method performs worse when the pathloss can take larger values, but
it can sometimes still give a good estimation of the jammer position. The method
of \cite{Borio-GNSS+-2016} performs a bit better than in previous tests. A 
higher value of $\alpha_i$ results in lower jamming levels at the receiver, 
sometimes low enough that the jamming is not detected if the receiver is too far
away. This actually helps method \cite{Borio-GNSS+-2016}, as the mean position 
is then computed based on receivers that are closer to the jammer. As a result, 
it estimates the jammer position better. However, this requires that the 
receivers are located close to the jammer, which might not always be the case.
For the proposed method, having receivers not detecting attacks or getting 
removed during the receiver selection due to a high $\alpha$, affects the 
algorithm negatively, because fewer receivers are used in the estimation. 
Testing an $\alpha$ that is just slightly higher than 2 would be of interest. 

\subsection{AGC}
\begin{figure}[t]
	\centering
	\includegraphics[width=1.0\columnwidth]{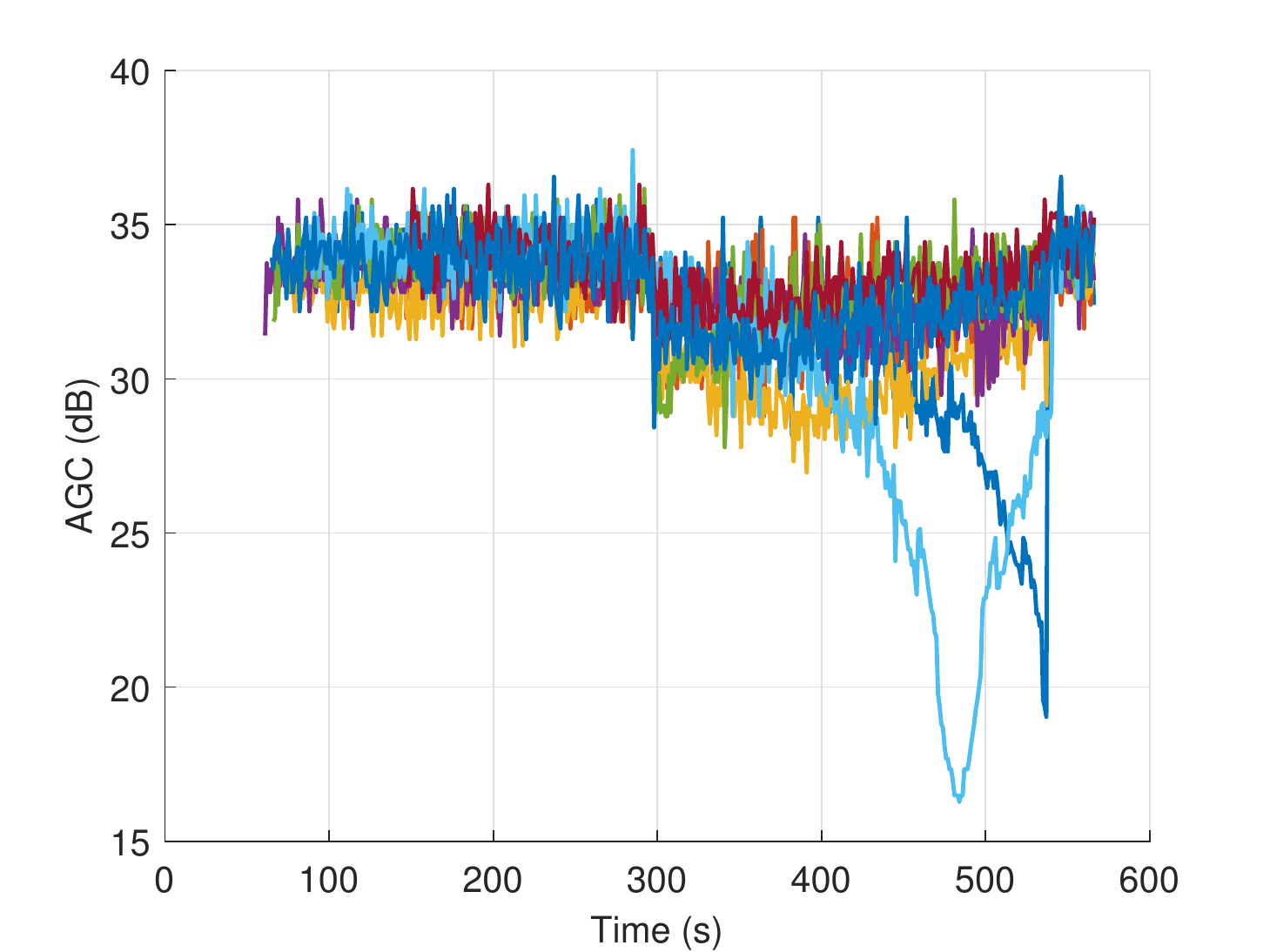}
	\caption{AGC for each receiver as reported by a Samsung S20+.}
	\label{fig:agc_levels}
\end{figure}

\begin{figure}[t]
	\centering
	\includegraphics[width=1.0\columnwidth]{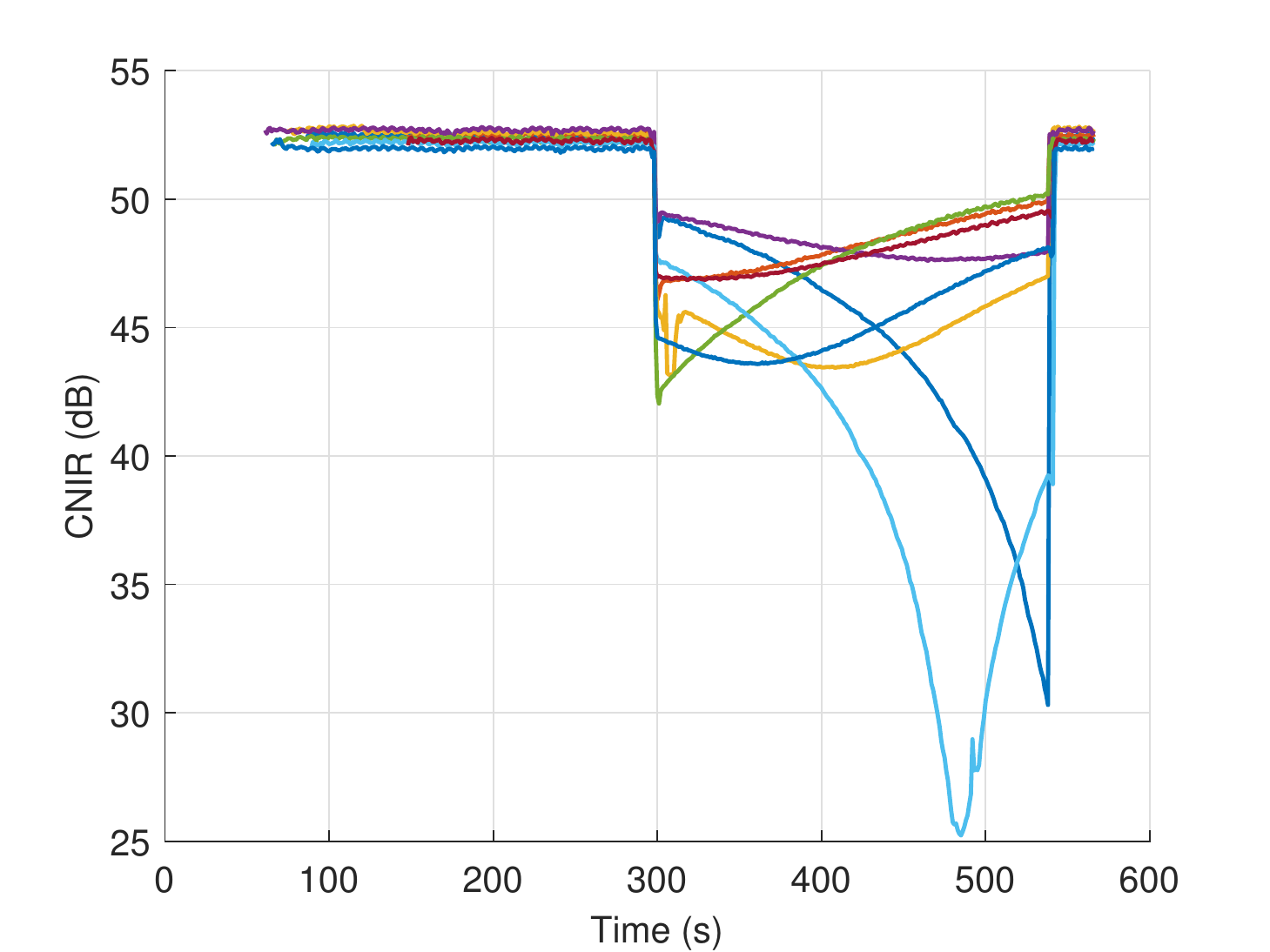}
	\caption{Average CNIR for each receiver as reported by a Samsung S20+.}
	\label{fig:cn0_levels}
\end{figure}

As stated in Section~\ref{sec:model}, AGC measurements can also be used with 
this algorithm in lieu of CNIR measurements. However, the raw AGC measurements 
output by the mobile phone used in our experiments are also affected by noise, 
to the extent that they do not provide any useful result at all. The AGC levels
from the eight simulated paths are shown in Figure \ref{fig:agc_levels}, as a 
function of simulation runtime. Such large fluctuations of AGC are not 
unreasonable, and similar variations for the same mobile phone model 
(Samsung S20+) were presented in \cite{Spens-NAVIGATION-2022}. This can be 
compared to the average CNIR estimates for each receiver, shown in Figure
\ref{fig:cn0_levels}, which are much smoother. However, the two figures show 
similar trends, so using a low-pass, median or a similar filter, might be enough
to smooth the data to a degree where it would be usable. This has not been done 
in this work, but it is within a possible topic for future work to investigate.

\section{Concluding Remarks}
\label{sec:conclusion}
We have extended the previous jammer position estimator 
\cite{Olsson-ICLGNSS-2022}by including received data only during parts of the 
sensing period when jamming is detected. Hardware tests using a Samsung S20+ 
mobile phone as the participatory sensors and a Spirent GSS9000 GNSS simulator 
to generate GNSS and jamming signals showed that the proposed algorithm works 
well for $C/N_0$ measurements and outperforms the compared state-of-the-art 
algorithms in the evaluated scenarios. Using 6 or more receivers gives a median 
error of about 50 meters, while the other methods give an error of 500 meters at 
best. When varying the pathloss, the error is larger, but the algorithm is still
able to find a solution. An $\alpha$ slightly higher than 2 should be tested in 
order to test the algorithm's capabilities to handle different pathlosses, but 
not so high that receivers do not detect the jamming or too many gets removed by
the receiver selection. The AGC output from the phone was too noisy and needs
further processing to be useful for position estimation. Such processing is not
done in this paper, but could be a topic for future work.


\end{document}